\newcommand{\cmark}{\textcolor{green!70!black}{\ding{51}}}%
\newcommand{\xmark}{\textcolor{red}{\ding{55}}}%

\documentclass[sigconf,nonacm]{acmart}
\usepackage{pifont}
\usepackage{booktabs}
\usepackage{multirow}
\AtBeginDocument{%
  }

\setcopyright{acmlicensed}
\copyrightyear{2026}
\acmYear{2026}
\acmConference[RecSys '26]{ACM Conference on Recommender Systems}{September 28--October 2,
  2026}{Minneapolis, MN, USA}

\begin{document}

\title{Reddit2Deezer: A Scalable Dataset for Real-World Grounded Conversational Music Recommendation}

\author{Haven Kim\quad Julian McAuley}
\email{{khaven, jmcauley}@ucsd.edu}
\affiliation{%
  \institution{University of California San Diego}
  \city{La Jolla, CA}
  \country{USA}}
\renewcommand{\shortauthors}{Kim and McAuley}

\begin{abstract}
Conversational music recommendation (CMR) research currently faces a tradeoff between authentic dialogue corpora that are limited in scale and synthesized corpora that scale up but whose conversations are artificially constructed rather than naturally observed. In this paper, we introduce \textbf{Reddit2Deezer}, a reality-grounded CMR resource derived from 190k unique \{thread, leaf-comment\} pairs. We release the resource in two versions: a raw version that preserves authenticity, and a paraphrased version that maximizes long-term reproducibility. Each musical entity is linked to a Deezer identifier, which provides straightforward access to audio previews and rich metadata (e.g., genre tags, popularity, BPM), opening the door to future research on content-grounded conversational recommendation. A human validation confirms the quality of the dialogues, item grounding, and paraphrases. The dataset is available at \url{https://huggingface.co/datasets/McAuley-Lab/Reddit2Deezer}.
\end{abstract}




\maketitle


\section{Introduction and Background}


Conversational recommendation, which extends classical recommenders by eliciting preferences through natural-language dialogue, has gained traction across domains such as travel destinations~\cite{goker2000adaptive, christakopoulou2016towards}, e-commerce~\cite{zhang2018towards}, movies~\cite{He_2023}, and music~\cite{doh2025talkplay, doh2025tools}.

One of the first human-collected conversational music recommendation (CMR) resources is CPCD~\cite{chaganty2023beyond}, which was collected by paid annotators. While foundational, its collection protocol inherently caps the achievable scale. To scale beyond this ceiling, a subsequent work~\cite{doh2024lpmusicdialog} ports CPCD's intent taxonomy into a GPT-3.5 pipeline grounded in the Million Song Dataset~\cite{bertin2011million}, whose catalog ends in 2010 and therefore excludes more than a decade of subsequent releases. An alternative approach~\cite{doh2025talkplay} replaces the explicit taxonomy with an LLM-based music-captioning model~\cite{doh2023lp} and an automatic tempo and chord recognition model~\cite{bock2016madmom}; however, captioning models can miss nuanced musical details or hallucinate and both components can make errors, and this noise can propagate into the synthesized dialogues. The most recent of these~\cite{choi2025talkplaydata2} proposes a four-agent (Profile / Goal / Listener / RecSys) pipeline built over LFM-2b~\cite{schedl2022lfm}; however, its topic distribution is hand-designed rather than derived from observed user behavior, and LFM-2b is no longer publicly available, constraining independent re-derivation. A parallel single-turn line of work~\cite{melchiorre2025just, palumbo2025text2tracks} trades multi-turn elicitation for scale by training on search and playlist logs, which are proprietary and thus cannot be redistributed. Closer to our approach, MusiCRS~\cite{surana2025musicrs} mines seven music subreddits and extracts entities and queries with open-source LLMs, following precedents from conversational movie recommendation~\cite{He_2023} and a related extension to e-commerce that maps items to Amazon products~\cite{jeon2025lavicadaptinglargevisionlanguage}. While this approach captures how people actually seek music recommendations in the wild, the dataset is limited in scale, comprising 477 conversations, with its items being grounded through YouTube links, which may become unavailable over time. Taken together, prior CMR resources reflect a tradeoff: the real-world-grounded ones are modest in scale, which limits their suitability for training modern models, while the larger synthetic ones may not fully reflect the distribution of real human music-seeking behavior. Compounding this, while music recommendation has historically leveraged an exceptionally rich set of signals—including audio, artist, genre tags, and popularity statistics~\cite{volkovs2018two, antenucci2018artist, doh2025talkplay, doh2025tools, kim2026fusid}, existing CMR resources are uneven in their support for these signals: some rely on grounding mechanisms that degrade over time (e.g., YouTube links) or on source corpora that are no longer accessible while some omit explicit linkage to audio or rich metadata.

\begin{table*}[t]
  \centering
  \caption{Conversational music recommendation datasets compared (as of May 2026).}
  \label{tab:cmr-comparison}
  \footnotesize
  \setlength{\tabcolsep}{3pt}
  \resizebox{0.99\textwidth}{!}{%
  \begin{tabular}{l l r r r r l l l l}
    \toprule
    Dataset
      &  \shortstack[r]{Public}
      & \shortstack[r]{\# Conv.}
      & \shortstack[r]{\# Multi-turn}
      & \# Tracks
      & \# Albums
      & Source corpus
      & Real / Synth.
      & Linked to
      & Catalog Cutoff \\
    \midrule
    CPCD~\cite{chaganty2023beyond}
      & \cmark
      & 917 & 917 & 107k & 0
      & Human raters & Real (human)
      & YouTube & $\sim$2022-2023 \\
    LP-MusicDialog~\cite{doh2024lpmusicdialog}
      & \cmark
      & 290k & 290k & 391k & 0
      & CPCD taxonomy + MSD~\cite{bertin2011million}, GPT-3.5 & Synthetic
      & MSD~\cite{bertin2011million} & 2010 \\
    TalkPlay~\cite{doh2025talkplay}
      & \xmark
      & 532k & 532k & 410k & 0
      & LP-MusicCaps~\cite{doh2023lp}, Madmom~\cite{bock2016madmom}, Whisper-Large-V3~\cite{radford2023robust} & Synthetic
      & Spotify MPD~\cite{chen2018recsys} & Oct. 2017 \\
    Text2Tracks~\cite{palumbo2025text2tracks}
      & \xmark
      & 1M & 0 & 500k & 0
      & Spotify playlists & Real (curated titles)
      & N/A & Not reported \\
    JAM~\cite{melchiorre2025just}
      & \xmark
      & 112k & 0 & 100k & 0
      & Deezer search logs & Real (search log)
      & N/A & 2024 \\
    TalkPlayData~2~\cite{choi2025talkplaydata2} & \cmark 
      & 16k & 16k &  47k & 0
      & LFM-2b, Gemini-2.5 Flash & Synthetic
      & LFM-2b + Spotify & 2020 \\
    MusiCRS~\cite{surana2025musicrs}
      & \cmark
      & 0.5k & 0.5k & \multicolumn{2}{c}{\hspace{1em} 3{,}589 (combined)}
      & Reddit (7 genre subreddits) & Real + LLM-augmented
      & YouTube & $\sim$2024-2025 \\
    \midrule
    \textbf{Reddit2Deezer (ours)}
      & \textbf{\cmark}
      & \textbf{190k} & \textbf{2k} & \textbf{100k}
      & \textbf{30k}
      & \textbf{Reddit (200 subreddits)}
      & \textbf{Real/Paraphrased}
      & \textbf{Deezer API} & \textbf{Apr. 2026} \\
    \bottomrule
  \end{tabular}%
  }
\end{table*}

To address these limitations jointly, we introduce \textbf{Reddit2Deezer}, a conversational music recommendation dataset built from real-world Reddit music-recommendation conversations, with musical entities linked to the Deezer API. This linkage enables straightforward access to audio previews and rich metadata (release date, track length, artist popularity, track and album popularity, BPM, and genre tags) without requiring an API key, and is far less prone to deletion than YouTube links. Because the conversations originate from ongoing Reddit discussions rather than legacy catalogs such as MSD~\cite{bertin2011million} or LFM-2b~\cite{schedl2022lfm}, recommendations naturally cover recent releases. Table~\ref{tab:cmr-comparison} contrasts our corpus with prior resources across several key factors; our dataset constitutes the largest real-world-grounded conversational music recommendation corpus, with a size comparable to that of synthetic ones, to our knowledge.


\section{Reddit2Deezer}

\subsection{Dataset Construction}

\noindent \textbf{Reddit Corpus Acquisition} \indent To obtain the widest tractable coverage, instead of relying on a hand-picked set of subreddits~\cite{surana2025musicrs}, we seed the subreddit list from the community-curated \texttt{r/Music/wiki/} \texttt{musicsubreddits} index—a long-running, human-edited registry of 695 music-related subreddits. We filter out subreddits that are unreachable at crawl time. Next, we manually inspect each remaining subreddit for topical relevance: many wiki-listed subreddits are dedicated to production, industry news, self-promotion, or music hardware rather than music discovery, and are therefore excluded. After both filters, 200 subreddits remain for the final crawl. Each retained subreddit is collected via the arctic-shift archive API over the period from January 2008 to mid-April 2026, as no subreddit predates January 2008. The upper bound reflects the crawl cut-off.

\noindent \textbf{Structural Filter} We apply an inexpensive structural pre-filter to discard records unlikely to contain a music discovery conversation before the LLM filtering stage, as running the LLM over the raw corpus—spanning 200 subreddits and 18 years—is prohibitive in GPU wall-clock time. Specifically, we first remove posts with no comments and comments whose parent post is missing, and then apply three content-level rules: (i) posts and comments whose body (after whitespace stripping) is under five characters are removed, adapting a heuristic from prior work~\cite{surana2025musicrs}; (ii) comments with a negative score are removed, as downvoted replies are unlikely to contain a helpful recommendation signal; and (iii) posts whose titles contain stopwords (e.g., \emph{favorite/favourite}, \emph{your opinion}) are removed, as such threads solicit broad subjective enumerations rather than targeted recommendations grounded in a stated need.

\noindent \textbf{LLM Filter} \indent To discard thread–comment pairs that are not music recommendation conversations and to automatically extract information about music entities, we use Qwen3.6-35B-A3B-FP8~\cite{yang2025qwen3technicalreport}, an LLM that ranks among the strongest open-source models on structured-extraction benchmarks~\cite{singh2026structuredoutputbenchmarkmultisource}. In the first stage, the model labels whether each thread is explicitly seeking music recommendation based on taste, mood, context, reference tracks, or criteria such as era or genre. We then apply the comment-filtering stage to comments from threads that pass this step by labeling each comment as either recommending a specific music item—along with extracted \{artist, title, type\}—or not, where the extracted information is later used for Deezer API linking. Both stages also output a self-reported confidence score in $[0,1]$ for each label, and we empirically use a threshold of 0.95 for downstream inclusion.

\noindent \textbf{Deduplication} \indent Reddit users often cross-post the same content both within a single subreddit and across related subreddits to enhance visibility. We therefore apply two deduplication processes. Within-subreddit duplicates are collapsed by normalized title and user within each subreddit; cross-subreddit duplicates are collapsed using the same approach after per-subreddit filtering. Both stages \emph{merge} rather than discard duplicates: comments are unioned, and each retained record stores the post IDs of the duplicates it absorbs (for both within-subreddit and cross-subreddit merges), ensuring that every retained record remains traceable to its original posts.

\noindent \textbf{Deezer API Linking} \indent To support downstream experiments (e.g., content-based conversational music recommendation), the grounding catalog must provide both metadata and audio. We chose Deezer because downloading audio previews does not require an API key unlike other platforms like Spotify, and Deezer identifiers are less prone to deletion than YouTube links. In addition, the API provides rich metadata, including popularity, BPM, and genre tags. To this end, we query Deezer with the artist and title extracted during the LLM filtering stage—when both the artist name and the title match (case-insensitive), we link the entity to the Deezer API.

\noindent \textbf{Paraphrasing} \indent To maximize long-term reproducibility, we provide a paraphrased version of the dataset alongside the raw version. To construct it, we ask Qwen3.6-35B-A3B-FP8~\cite{yang2025qwen3technicalreport} to paraphrase each \{\texttt{thread\_id}, \texttt{leaf\_id}\} pair according to the following rules: preserve every music-relevant detail verbatim without altering artist names or track/album titles; resolve relative time references using the post year (e.g., ``in 2025'' rather than ``this year''); strip or substitute overt personal information; and restyle the exchange as a one-on-one music recommendation chat (e.g., ``Hi'' rather than ``hey reddit''). The prompt includes the aforementioned rules and nine human-written examples, totaling 10,336 words; the full prompt will be released upon acceptance.


\subsection{Human Validation}

Because our dataset is automatically constructed, we conduct a human validation study to verify (i) whether the resulting conversations are genuine music recommendation dialogues, (ii) whether the Deezer identifiers are accurately mapped to the recommended songs, and (iii) whether the paraphrased versions preserve the musical details~\cite{doh2024lpmusicdialog}. To determine an appropriate sample size, one author first conducted a pilot screening of 100 random samples from the raw data, identifying three negative cases in (i) and two negative cases in (ii). On this basis, we adopted an expected prevalence of $p=0.05$ and applied Cochran's formula~\cite{cochran1963sampling} ($n = \frac{Z^2 p (1-p)}{e^2}$) with a 95\% confidence level ($Z=1.96$) and an absolute margin of error of $e=0.05$, yielding a minimum sample size of 73.

Before conducting the survey, we provided participants with a definition of a music recommendation conversation—namely, an exchange in which the seeker is soliciting music recommendations and the recommender is recommending music—along with two positive and two negative examples. For the faithfulness rating, we additionally showed five \{rating, raw–paraphrased pair\} examples, each accompanied by a brief rationale. All examples provided to participants were drawn from the earlier pilot screening. Two non-author participants then independently annotated 73 randomly sampled raw–paraphrased pairs, matched by \{thread\_id, leaf\_id\}. For each pair, annotators provided three labels: a binary label indicating whether the conversation is a genuine music recommendation conversation, a binary label indicating whether the Deezer identifier is accurately mapped to the recommended track/album, and a faithfulness rating indicating the degree to which the paraphrased version preserves the musical details (0–5 scale, where higher scores indicate higher fidelity).

(i) The two annotators identified 71 and 72 out of 73 raw conversations, respectively, as valid music recommendation dialogues, and all 73 paraphrased conversations as valid. The inter-annotator agreement rate is 99.3\%, with Cohen's $\kappa = 0.66$~\cite{feinstein1990high}. The near-perfect validity of the paraphrased version is attributable to the fact that our paraphrasing prompt explicitly instructs Qwen to restyle each exchange as a one-on-one music recommendation chat. (ii) The annotators judged 93.84\% of the Deezer identifier mappings to be correct on average, with an agreement rate of 95.89\% and $\kappa = 0.65$, in both versions, as none of the 73 paraphrased samples recommended a different artist or title from its raw counterpart. The mismatches mostly stem from version differences (e.g., remix, remastered) and disagreements typically arose when one annotator marked a live or instrumental version as a mismatch while the other accepted it as correct. (iii) The mean ratings from two annotators for how musically faithful the paraphrased version is to the original were 4.84 ± 0.21 and 4.82 ± 0.33 (on a 0–5 scale), with a Quadratic Weighted Kappa of 0.30. Overall, these results indicate that the dataset consists primarily of valid music recommendation conversations with canonical Deezer identifiers, and that the paraphrased version is musically faithful to the original text.
\begin{table}[t]
\centering
\caption{Corpus statistics for the Reddit2Deezer dataset.
         $\Delta$\,\% = (paraphrased $-$ raw) / raw $\times$ 100.}
\label{tab:corpus-stats}
\resizebox{0.80\linewidth}{!}{%
\begin{tabular}{lrrr}
\toprule
\textbf{Metric} & \textbf{Raw} & \textbf{Paraphrased} & $\boldsymbol{\Delta}$\,\textbf{\%} \\
\midrule
Conversations                       & 186{,}380 & 535{,}592 & $+$187.3 \\
Single-turn         & 184{,}048 & 524{,}531 & $+$185.0 \\
Unique thread IDs                   &  41{,}286 &  41{,}251 & $-$0.1   \\
Unique (thread, leaf)       & 186{,}380 & 185{,}720 & $-$0.4   \\
\midrule
Avg.\ rec.\ turns.         &     1.01 &     1.02 & $+$0.8   \\
Avg.\ items per turn          &     1.51 &     1.00 & $-$33.7  \\
\midrule
Unique artists                      &  42{,}497 &  42{,}393 & $-$0.2   \\
Unique track IDs                    & 100{,}832 & 100{,}439 & $-$0.4   \\
Unique album IDs                    &  29,178 &   28,911 & $-$0.9   \\
\bottomrule
\end{tabular}%
}
\end{table}

\subsection{Corpus Statistics}

Table~\ref{tab:corpus-stats} summarizes the statistical comparison between the two versions. Five patterns are worth highlighting. (i) The small residual loss in unique-prompt and unique-item counts is observed in the paraphrased data, and it is attributable to cases where the LLM occasionally fails to emit a parsable artist and title, causing the record to be discarded. (ii) The conversation count nearly triples in the paraphrased version. This is because, rather than emitting one paraphrase per source comment, we emit one paraphrase per \{thread, leaf-comment, verified-item\} triple, so each paraphrased dialogue focuses on a single recommendation per turn. (iii) As a direct consequence of (ii), the paraphrased version averages 1.0 items per recommender turn, whereas the raw version averages 1.5 with implications for set-valued predictions rather than single-item predictions. (iv) Reflecting real-world music-seeking behavior, a non-trivial share of recommendations are at the album level rather than the track level: our dataset includes roughly 30k unique albums in addition to 100k unique tracks, whereas existing CMR datasets are almost exclusively track-only (Table~\ref{tab:cmr-comparison}). (v) Our dataset is single-turn-dominant, with only about 2k multi-turn conversations; nevertheless, we show in Section~\ref{sec:results} that this single-turn corpus can still be leveraged to improve multi-turn conversational music recommendation.


\begin{table*}[]
\centering
\caption{Test-set recommendation performance. \textbf{H}, \textbf{R}, and
           \textbf{N} denote Hit@$k$, Recall@$k$, and NDCG@$k$
           respectively. \emph{All values are multiplied by 100.} Best
           value within each column half is bolded.}
  \label{tab:results}
  \resizebox{0.99\textwidth}{!}{%
\begin{tabular}{@{}llrrrrrrrrrrrrrrrrrr@{}}
\toprule
& & \multicolumn{9}{c}{\textbf{Raw test set}} & \multicolumn{9}{c}{\textbf{Paraphrased test set}} \\
\cmidrule(lr){3-11} \cmidrule(lr){12-20}
\textbf{Backbone} & \textbf{Variant}
& \textbf{H@1} & \textbf{R@1} & \textbf{N@1}
& \textbf{H@5} & \textbf{R@5} & \textbf{N@5}
& \textbf{H@20} & \textbf{R@20} & \textbf{N@20}
& \textbf{H@1} & \textbf{R@1} & \textbf{N@1}
& \textbf{H@5} & \textbf{R@5} & \textbf{N@5}
& \textbf{H@20} & \textbf{R@20} & \textbf{N@20} \\
\midrule
\multirow{2}{*}{\textbf{Retrieval}}
& CLAP-Audio & 0.15 & 0.04 & 0.15 & 0.23 & 0.04 & 0.08 & 0.62 & 0.08 & 0.09 & 0.00 & 0.00 & 0.00 & 0.23 & 0.11 & 0.06 & 1.08 & 0.60 & 0.14 \\
& CLAP-Text  & 0.00 & 0.00 & 0.00 & 0.00 & 0.00 & 0.00 & 0.31 & 0.03 & 0.02 & 0.00 & 0.00 & 0.00 & 0.23 & 0.01 & 0.03 & 0.85 & 0.08 & 0.06 \\
\midrule
\multirow{3}{*}{\textbf{Generative}}
& Zero-shot & 0.31 & 0.10 & 0.31 & 1.16 & 0.47 & 0.42 & 2.39 & 0.65 & 0.52 & 0.93 & 0.27 & 0.93 & 1.93 & 0.61 & 0.77 & 3.32 & 0.90 & 0.82 \\
& FT-Raw    & \textbf{3.55} & \textbf{0.76} & \textbf{3.55} & \textbf{9.18} & \textbf{2.12} & \textbf{3.01} & \textbf{17.67} & \textbf{3.93} & \textbf{3.23} & 1.93 & 0.35 & 1.93 & 6.79 & 1.31 & 1.94 & 12.73 & 2.72 & 2.09 \\
& FT-Para   & 2.31 & 0.34 & 2.31 & 7.41 & 1.64 & 2.27 & 14.20 & 3.26 & 2.50 & \textbf{3.32} & \textbf{0.68} & \textbf{3.32} & \textbf{9.95} & \textbf{2.25} & \textbf{3.08} & \textbf{17.36} & \textbf{4.00} & \textbf{3.28} \\
\bottomrule
\end{tabular}
}
\end{table*}


\section{Experiments}

The goals of this section are two-fold: to demonstrate concrete usage patterns on Reddit2Deezer for downstream conversational music recommendation modelling, and to compare the relative utility of the raw and paraphrased versions. Every model is evaluated on both versions' test partitions, so that the effects of train- and test-time text style can be disentangled. In our experiments, we use two families of models (retrieval and generative) to retrieve tracks or albums from the full catalog, comprising 130,010 items (100,832 tracks and 29,178 albums).

\noindent \textbf{Retrieval Models (CLAP-based)} \indent We embed both the conversation prior to the recommendation and every catalog item with a pre-trained text-audio joint encoder~\cite{wu2023large}, similar to the prior work~\cite{surana2025musicrs}. We represent each catalog item using two types of embeddings—audio and text—yielding two retrieval variants, which we denote \textbf{CLAP-Audio} and \textbf{CLAP-Text} respectively. For CLAP-Audio, we extract audio embedding from each item's 30-seconds Deezer preview by chunking and pooling over $3 \times 10$\,s windows, then re-normalizing. For CLAP-Text, we use Deezer metadata to render every catalog item as a structured natural-language description, covering artist, title, release date, duration, BPM, gain, explicit-lyrics flag, and popularity tiers. Because Deezer provides popularity-related information numerically and CLAP is unlikely to interpret relative popularity from raw numbers, we bucket these fields, with each bucket boundary set at one decade on the underlying integer scale: track popularity is bucketed from Deezer's rank field into \{viral, hit, well-known, moderate, deep cut, obscure\}, and artist popularity is bucketed from Deezer's number of fans into \{iconic, mainstream, well-known, established, underground, obscure\}. At inference, both variants retrieve tracks or albums by ranking catalog items according to the cosine similarity between their embeddings and the conversation embedding.

\noindent \textbf{Generative Models (Qwen3.5-based)} \indent In the \textbf{zero-shot} approach, we prompt Qwen3.5-2B~\cite{yang2025qwen3technicalreport} without any further training and instruct it to recommend one musical entity as a JSON object of the form \verb|{"artist", "title"}|. Predictions are matched against the catalog by case-insensitive artist-title lookup. In the fine-tuning approaches, we fine-tune the same model to output the same JSON artist-title format conditioned on the context. We split the dataset chronologically, following prior work~\cite{doh2025talkplay}, with a target ratio of approximately 90:5:5 (train:val:test): conversations before August 2025, between August 2025 and January 2026, and after January 2026 form the training, validation, and test sets, respectively. The two fine-tuning settings differ only in the training and validation corpus: \textbf{FT-Raw} is trained and validated on raw conversations, and \textbf{FT-Para} is trained on paraphrased conversations. For threads with multiple recommendations, the supervision target is sampled uniformly each epoch so that all candidate items are seen over the course of training. During validation and evaluation, on the other hand, all recommendations associated with a given thread are treated as valid answers. To preserve the structural cues of a Reddit thread, we add three special tokens to the tokenizer when training on raw data: \texttt{<subreddit>}, \texttt{<title>}, and \texttt{<body>}; we omit them when training and evaluating on paraphrased data because the paraphrased rewrites are self-contained without a separate post title and body. For both fine-tuned settings, we use AdamW~\cite{loshchilov2019decoupledweightdecayregularization} with learning rate $2e-4$~\cite{hu2021loralowrankadaptationlarge}, and a cosine schedule with $3\%$ linear warmup. The effective batch size is $256$ and the maximum sequence length is set to $512$. We fine-tune with LoRA adapters~\cite{hu2021loralowrankadaptationlarge} (rank $r{=}16$, $\alpha{=}32$, dropout $0.05$), while the new tokens' embedding and LM-head rows are unfrozen. Finally, across the three approaches, recommendations not in the catalog are discarded. Following prior work~\cite{ju2025handbook}, we stop training once the performance on the validation set has not improved over 10 validation intervals (100 steps each).


\section{Results}\label{sec:results}
Following prior conversational music recommendation works, we report results using Hit@$k$~\cite{doh2025talkplay, doh2025tools}, Recall@$k$~\cite{surana2025musicrs}, and nDCG@$k$~\cite{surana2025musicrs} with $k\in\{1,5,20\}$~\cite{doh2025tools, surana2025musicrs}. 

\noindent \textbf{Overall Performance} \indent Table~\ref{tab:results} summarizes performance on both the raw and paraphrased test partitions; within the CLAP-based retrieval family, CLAP-Audio outperforms the CLAP-Text across nearly every $(k,\text{test set})$. This gap suggests that representations derived directly from acoustic content provide more discriminative signal for matching seeker queries to catalog items than representations derived from explicitly encoding artist, title, era, and popularity tiers. Both retrieval variants also generally perform better on the paraphrased test set than on the raw one. We attribute this gap to the tendency where raw Reddit dialogues contain platform-specific artifacts (e.g.,  ``hey r/jazz,'') that lie outside the text distribution CLAP was pre-trained on, whereas the paraphrased version restyles each exchange as a clean one-on-one music conversation with platform-specific artifacts cleared. On the other hand, despite being a relatively small model and receiving no task-specific training, zero-shot Qwen3.5-2B surpasses both retrieval variants by a wide margin on every metric. This indicates that pre-trained LLMs already encode a non-trivial amount of musical knowledge—of artists, genres, and stylistic associations—relevant for conversational music recommendation, even before any in-domain supervision.

As for the fine-tuned variants, as expected, FT-Raw is the strongest model on the raw test set and FT-Para is the strongest on the paraphrased test set. At $k\in\{5,20\}$, FT-Para on the paraphrased set generally outperforms FT-Raw on the raw set. At $k=1$, however, FT-Raw on the raw set outperforms FT-Para on the paraphrased set. This is attributable to the fact that FT-Raw observes a slightly larger pool of unique supervision targets during training, since paraphrasing occasionally fails to emit a parsable artist-title pair (Section~2.3), and this catalog-coverage advantage matters most at $k=1$ where the right answer must be the top-1 prediction. For example, if a thread has two valid recommendations $\{A, B\}$ but $B$'s artist-title pair fails to parse during paraphrasing, FT-Raw can predict either $A$ or $B$ and still hit the top-1 answer, whereas FT-Para has only ever seen $A$ as a target. Both fine-tuned variants also transfer to their counterpart partition to some extent; however, FT-Para transfers to the raw set more effectively than FT-Raw transfers to the paraphrased set. We attribute this asymmetry to a distributional property of the two training sets: the paraphrased set provides a cleaner, platform-agnostic view of the task, so FT-Para learns more transferable representations of music-seeking dialogue, whereas FT-Raw partly entangles its learned signal with platform-specific structural and stylistic cues.

\begin{figure}[!t]
\centering
\includegraphics[width=0.60\columnwidth]{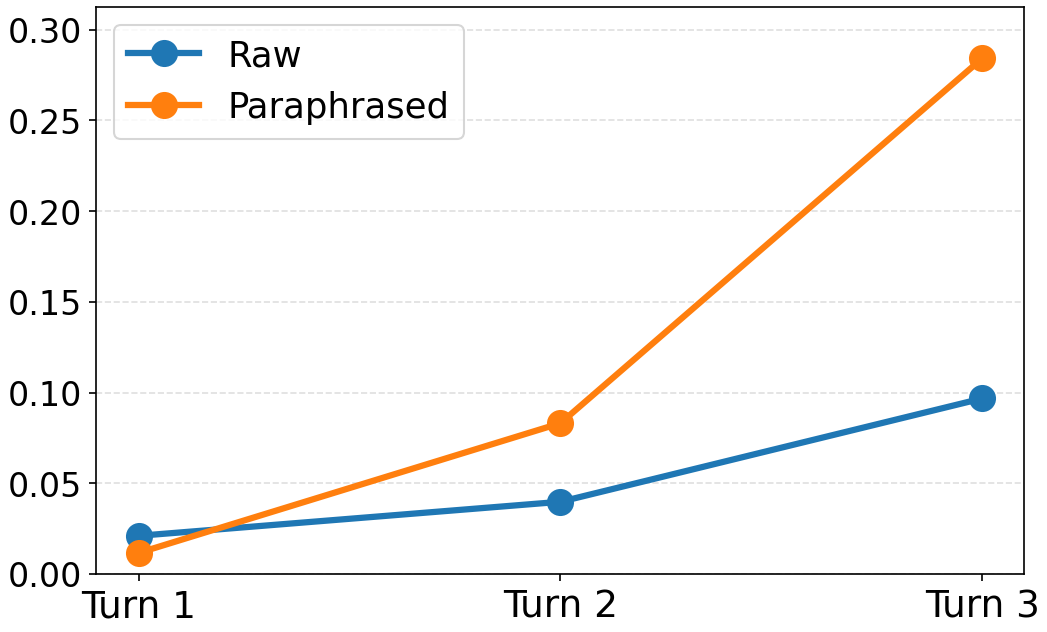}
\caption{nDCG@5 by recommender turn position. Per-turn sample sizes: $1220 / 81 / 2$ at Turns $1 / 2 / 3$. }
\label{fig:over-turn}
\end{figure}

\noindent \textbf{Performance over Turns} \indent Figure~\ref{fig:over-turn} plots nDCG@5 against recommender turn position for FT-Raw on the raw test set and FT-Para on the paraphrased test set. Although Reddit2Deezer is dominated by single-turn conversations (Table~\ref{tab:corpus-stats}), models fine-tuned on it nevertheless improve as conversations extend over multiple turns, indicating that the additional preference signal is exploited even though such trajectories are rare in training. Overall, this trend suggests that the model learns to narrow down the seeker's preferences as more dialogue context accumulates, supporting the use of Reddit2Deezer as a training resource for conversational music recommendation despite its single-turn-dominant composition. Nonetheless, we note that Turn 3 estimates are based on only two conversations and are therefore noisy.


\section{Conclusion}

We introduced Reddit2Deezer, a scalable CMR dataset that pairs music recommendation Reddit-sourced dialogues with Deezer item identifiers. The dataset is released in two versions: a raw version that maximizes authenticity and a paraphrased version that maximizes long-term reproducibility, the majority of both confirmed to be music discovery dialogues by human verifiers. To our knowledge, this is the largest reality-grounded conversational music recommendation dataset to date. Because Deezer item identifiers allow audio previews to be easily re-fetched via the public API—and we additionally release CLAP-audio embeddings—and because the Deezer API also provides access to rich metadata, building a conversational music recommender system on top of these resources is a natural direction for follow-up work.

\section{Acknowledgements}
This work is partially supported by NSF IIS-2432486.


\bibliographystyle{ACM-Reference-Format}
\bibliography{sample-base}

\end{document}